\documentclass{article}

\usepackage{arxiv2}

\usepackage[utf8]{inputenc} 
\usepackage[T1]{fontenc}    
\usepackage{hyperref}       
\usepackage{url}            
\usepackage{booktabs}       
\usepackage{amsfonts}       
\usepackage{nicefrac}       
\usepackage{microtype}      
\usepackage[utf8]{inputenc}
\usepackage{graphicx}
\usepackage{natbib}
\usepackage{doi}
\usepackage{authblk}
\usepackage{subcaption}
\usepackage{multirow}
\usepackage{bm}
\usepackage{siunitx}
\usepackage{soul}

\title{The rise of data-driven weather forecasting\\
\large{ a first statistical assessment of machine learning-based weather forecasts\\ in an operational-like context}}

\date{}
\author{Zied Ben Bouall\`egue, Mariana C A Clare, Linus Magnusson, Estibaliz Gasc\'on, Michael Maier-Gerber, Martin Janou\v{s}ek, Mark Rodwell, Florian Pinault, Jesper S Dramsch, Simon T K Lang, Baudouin Raoult, Florence Rabier, Matthieu Chevallier, Irina Sandu, Peter Dueben, Matthew Chantry, Florian Pappenberger}
\affil{ECMWF}
\begin{document}

\maketitle

\begin{abstract}
Data-driven modeling based on machine learning (ML) is showing enormous potential for weather forecasting. Rapid progress has been made with impressive results for some applications. The uptake of ML methods could be a game-changer for the incremental progress in traditional numerical weather prediction (NWP) known as the “quiet revolution” of weather forecasting. 
The computational cost of running a forecast with standard NWP systems greatly hinders the improvements that can be made from increasing model resolution and ensemble sizes. An emerging new generation of ML models, developed using high-quality reanalysis datasets like ERA5 for training, allow forecasts that require much lower computational costs and that are highly-competitive 
in terms of accuracy.  Here, we compare for the first time ML-generated forecasts with standard NWP-based forecasts in an operational-like context, initialized from the same initial conditions. Focusing on deterministic forecasts, we apply common forecast verification tools to assess to what extent a data-driven forecast  produced with one of the recently developed ML models (PanguWeather) matches the quality and attributes of a forecast from one of the leading global NWP systems (the ECMWF IFS). The results are very promising, with comparable accuracy for both global metrics and extreme events, when verified against both the operational IFS analysis and synoptic observations. 
Overly smooth forecasts, increasing bias with forecast lead time,
and poor performance in predicting tropical cyclone intensity are identified as current drawbacks of ML-based forecasts. A new NWP paradigm is emerging 
relying on inference from ML models and state-of-the-art analysis and reanalysis datasets for forecast initialization and model training. 
\end{abstract}

\section{Introduction}\label{sec:intro}

Numerical weather prediction (NWP) is the dominant approach for weather forecasting. A weather forecast is the result of the numerical integration of partial differential equations starting from the best estimate of the current state of the Earth System. The idea that the physical laws of fluid dynamics and thermodynamics can be used to predict the state of the atmosphere dates back to the pioneering works of \cite{abbe1901} and \cite{bjerknes1904}. In a standard NWP framework, a weather prediction results from a deductive inference: a deterministic forecast is derived using the laws of physics starting from the best possible initial conditions, derived by optimally combining earth system observations and short-range forecasts through data assimilation. However, our ability to perfectly know the initial conditions and numerically resolve the equations is limited. Hence, ensemble forecasting is used to account for uncertainty in both the initial conditions and the forecasting model, with the resulting ensemble forecast serving as a basis for probabilistic forecasting \citep{leutbecher2008}.

\begin{figure}
\centering
\includegraphics[width = 0.95\textwidth,trim={0cm 0cm 0cm 0cm},clip]{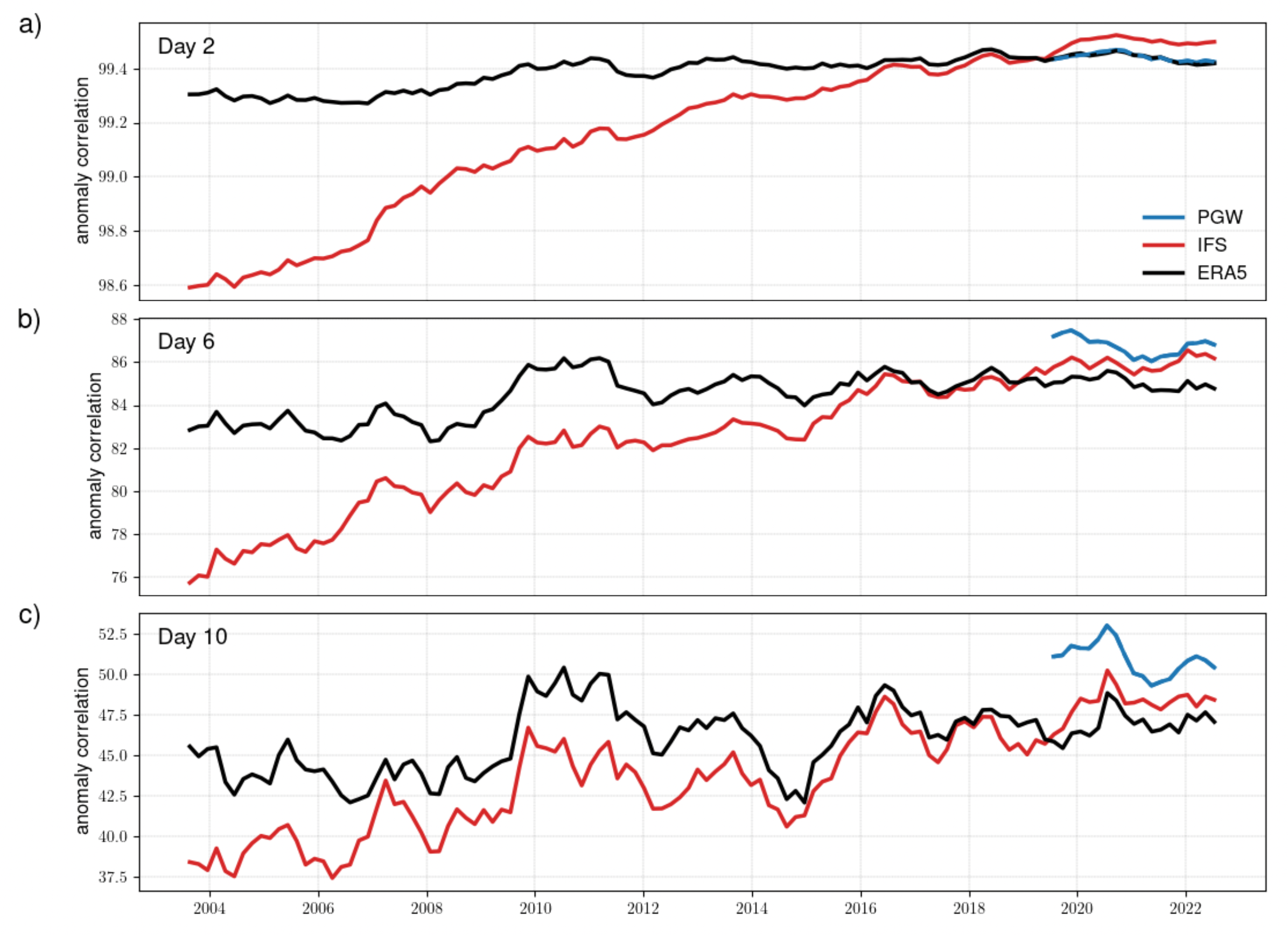}
\caption{Forecast accuracy (the larger the better) over the Northern Hemisphere at day 2 (a), day 6 (b), and day 10 (c). Accuracy is measured as the correlation between the forecasts and the verifying analysis for the geopotential height at 500-hPa, expressed as the anomaly with respect to the climatological height. A one-year running mean is applied. The constant improvement over the past decades of the IFS forecasts is compared with the performance of the ERA5 forecast (run with the IFS version operational in 2016) and with the performance of the PGW forecast trained over 1979-2018 and verified over 2019-2023.}
\label{fig:hsc}
\end{figure}

A continuous improvement of the NWP performance has been observed over the last decades, including for the prediction of high-impact weather events \citep{zbb19}. Skill improvement is achieved through improvements in initial conditions, numerical models, and resolution. At the European Centre for Medium-Range Weather Forecasting (ECMWF), the Integrated Forecasting System (IFS) has been run operationally since 1979 with regular updates of the different components of the forecasting system. The evolution of the IFS accuracy over the last two decades is shown in Fig.~\ref{fig:hsc} (red lines). The steady increase in forecast accuracy thanks to incremental improvements in numerical modeling, supercomputing, data assimilation and ensemble techniques, observations and their use in the NWP system, has become known as the `quiet revolution' of weather forecasting \citep{bauer2015}. 
However, the computational cost of running a forecast is a major bottleneck that hinders rapid improvements with standard NWP systems. In operational NWP, the computational and timeliness constraints imply finding a balance between increasing model resolution and increasing ensemble size, which are two major factors known to improve the skill of ensemble forecasts \citep{leutbecher2020}.  

In recent years, data-driven modeling based on machine learning (ML) is showing large potential for weather forecasting applications with the promise to deliver forecasts at a much lower computational cost along with potential other benefits such as increased timeliness and potentially increased accuracy \citep{burgh2023}. Pioneer works used simple convolutional-based neural networks to predict a small subset of variables using only these variables as predictors \citep[][]{dueben2018, weyn2019}. Further developments employed more complex neural networks and more variables as predictors resulting in more accurate machine-learning models that were nevertheless still considerably less accurate than NWP systems \citep[][]{weyn2020improving,rasp2021data}. However, since 2022, tremendous progress has been made with a series of key works developing machine-learning models for weather forecasting, presenting impressive forecast scores for a large number of different weather variables, some of which rival the operational ECMWF deterministic high-resolution (deterministic) forecasts \citep{keisler2022forecasting, pathak2022fourcastnet, bi2022pangu, lam2022graphcast, chen2023fengwu}. Concretely, \cite{keisler2022forecasting} use a Graph Neural Network (GNN) model and claim to produce more accurate forecasts of specific humidity than IFS after day 3; \cite{pathak2022fourcastnet} leverage Fourier Transforms with a transformer and claim to produce comparable accuracy to IFS for 2m temperature; \cite{bi2022pangu} use a vision transformer model and claim to produce more accurate forecasts than IFS across numerous variables when both models are verified against reanalysis;  \cite{lam2022graphcast} use a GNN and claim more accurate forecasts than IFS on a larger set of atmospheric variables and pressure levels; and finally \cite{chen2023fengwu} use a transformer and claim to improve the scores compared to \cite{lam2022graphcast}, especially at longer lead times. 

The emergence of data-driven models has been made possible thanks to the 
availability of large, high-quality, open, and free meteorological datasets.
The aforementioned ML models are trained on ERA5 reanalysis data, which is the fifth generation ECMWF atmospheric reanalysis, produced by the Copernicus Climate Change service, as one of the European Union Copernicus Programme key deliverables \citep{era5}. This dataset is particularly attractive for machine learning problems because it is a continuous weather dataset from 1940 to the present day and it represents the best possible reconstruction of the Earth-system state created by blending past observations and short-range forecasts through data assimilation. However, the ML methods presented only train from 1979 because the extensions to 1940 are relatively recent and have lower accuracy due to the very limited availability of satellite data before 1980 \cite[e.g.][]{era5extension}. ERA5 is generated using the operational IFS cycle at the time of production (2016) and is publicly available at a grid resolution of $0.25^{\circ}$ (28km). Hence, ML models are trained on reanalysis at a much lower resolution than that of today's operational forecasts and analyses (28km instead of 9km in the case of the ECMWF operational high-resolution forecasts and analyses). Note that, despite this resolution difference, ERA5 ``forecasts'' are used routinely for forecast verification purposes: the performance of the current IFS is compared with the performance of ERA5 forecasts (10-day forecast initialized from ERA5 at the ERA5 resolution, about 25 km) to help distinguish inter-annual variability from actual skill improvement due to changes in the forecasting system. This is illustrated in Fig.~\ref{fig:hsc} with ERA5 forecast accuracy represented by black lines. 

One approach to data-driven weather prediction would consist of running ML-trained models starting from optimized initial conditions in an operational context. In such a 
weather prediction system, the forecast inference relies on an ML model rather than the physical model (included for example in the IFS). This approach is highly attractive because a forecast can be generated at a speed several orders of magnitude faster than that from conventional methods. At a fundamental level, an ML-based prediction is the result of an inductive rather than a deductive inference. This paradigm shift in terms of logic has implications for the way a weather forecast is interpreted: a forecast becomes a plausible outcome given what has been learned from previous data. However, the mode of inference followed by ML methods can raise concerns, in particular regarding the ability of such models to predict extreme events unseen in the training dataset. Moreover, the interpretability of ML models is also often questioned when they are perceived as black boxes where the link between the training dataset and the current forecast is difficult to grasp \citep{mcgovern2019}. The huge potential benefits and drawbacks of data-driven systems trigger the question of whether ML models can become a component of operational NWP systems.

In this study, we evaluate the performance of data-driven forecasts in an operational-like context. More precisely, the PanguWeather ML model in \cite{bi2022pangu} (referred to hereafter as PGW), which is open-source for non-commercial use, has been set up to run on the ECMWF computers. For the first time, a forecast generated with an ML model is compared with an operational NWP forecast using the same framework and starting from the same initial conditions. In \cite{bi2022pangu}, like in previous studies, the ML-based forecasts were initialized from ERA5. We leverage standard verification techniques routinely applied for weather forecast evaluation at ECMWF. Using this methodology, we can assess which aspects of the data-driven forecasts can match the quality of forecasts performed with one of the leading operational NWP systems. This study focuses predominantly on the statistical analysis of forecast performance, but we acknowledge that case studies play a key role in understanding the capability and limitations of ML models in weather forecasting and refer the reader to \cite{magnussonNL2023}.
Also, the practical value of these forecasts would need to be carefully assessed in partnership with experienced forecasters, as discussed in \cite{Uphoff2023}.
 
\section{Methodology and experiments}
\label{sec:methodology}

Our comparative work is based on implementing PGW in an operational-like setting. PGW uses a vision transformer model architecture \citep{dosovitskiy2020image} with 3D weather fields as inputs and outputs. Developed in \cite{bi2022pangu}, the network minimizes a loss function defined as the root mean square error (RMSE) with a cos-latitude weighting to account for the spherical nature of the Earth as the model is trained on a regular latitude-longitude grid. Like most ML models, PGW uses an iterative method to forecast forward in time. A novelty of their approach, however, is that they choose to minimize the RMSE over a series of fixed short time periods (1hr, 3hrs, 6hrs, and 24hrs) and then achieve weather forecasts at any time using a Hierarchical Temporal Aggregation method. Here, PGW is run for 10 days, using a model time step of 24h. Though PGW demonstrated reasonable results beyond this time period, only the first 10 days are analyzed here.

For the verification periods under focus in this work, IFS is run operationally at a horizontal grid resolution of 9km up to 10 days lead time using the operational IFS cycle at the time (43r3 and 45r1 for 2018, 47r3 for 2022). We also include the publicly available ERA5 forecast in our comparison. 
ERA5 forecasts start from ERA5 reanalysis and are based on a lower model resolution than the operational IFS forecast (30km instead of 9km). 
Here we recall that the ERA5 reanalysis and forecasts are produced with a similar set-up of IFS as that used for the operational high-resolution forecasts and analysis, but they are produced with the operational cycle at the time when the production of ERA5 started (41r2, in 2016), and at a lower resolution (30km instead of 9km). Thus, ERA5 forecasts and IFS forecasts differ in terms of initial conditions, resolution, and IFS cycle.  

We choose to initialize PGW from the same analysis as IFS, namely the ECMWF operational IFS analysis. This choice appears natural for a fair comparison between PGW and IFS in an operational-like setting.  PGW was trained using ERA5 data and has a horizontal grid spacing of 28km, meaning initialization from the operational IFS analysis may induce some impact on scores. An optimal configuration of an ML-based forecasting system would likely `fine-tune' (training near convergence) on operational IFS analysis. This optimization is outside the scope of this work but is worthy of note. The operational IFS analysis is created by using the current operational data assimilation system, which operates at a higher resolution (9km) and uses a more recent (and therefore improved) IFS version than ERA5 to construct superior initial conditions.

\begin{figure}
\begin{center}
\includegraphics[width = 0.95\textwidth,trim={0cm 0cm 0cm 0cm},clip]{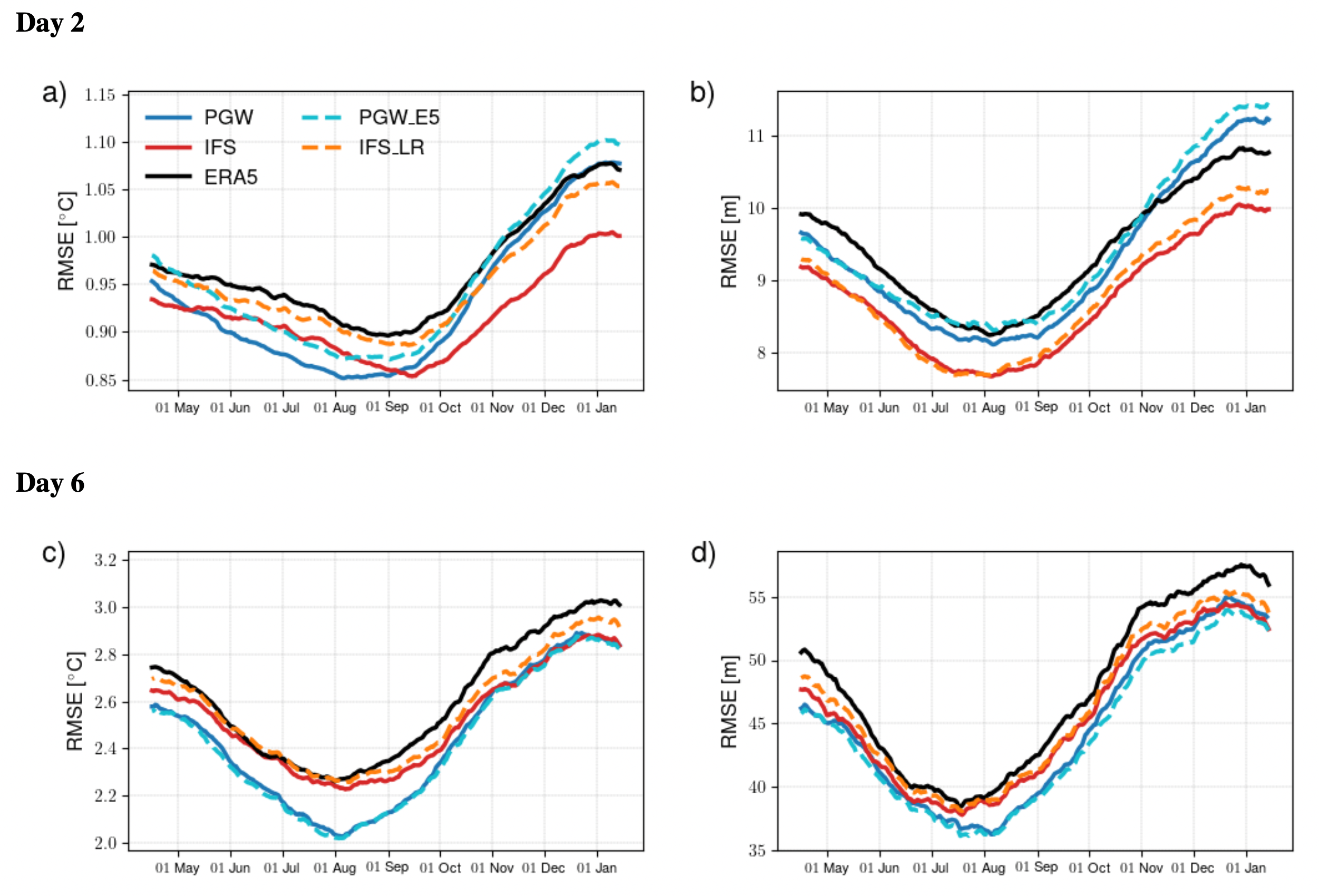}
\end{center}
\caption{The seasonal cycle of the RMSE (3-month averaged) at day 2 and day 6 over one year covering the period 01 March 2022 to 28 February 2023 aggregated over the Northern Hemisphere for  (a,c) T850 and (b,d) Z500. PGW\_E5: PGW initialized with ERA5, IFS\_LR: IFS run at a grid-resolution of 0.25$^\circ$. The forecasts are verified against operational IFS analyses on a grid resolution of 1.5$^\circ$. }
\label{fig:seasons}
\end{figure}

As a complementary experiment, we also run PGW starting from ERA5 reanalysis (PGW\_E5). This experiment shows that PGW starting from the operational high-resolution analysis generally performs better than PGW starting from ERA5 for the first days of the forecast (roughly up to day 4 depending on the variable and domain of interest), as illustrated in Fig.~\ref{fig:seasons}.

We also run IFS initialized from the operational IFS analysis at a lower resolution (IFS\_LR) close to the PGW resolution to isolate the impact of model horizontal resolution on forecast performance. This impact differs depending on both variable and lead time. For T850 at a lead time of two days, a change in horizontal resolution results in a clear degradation of the forecast accuracy, but at a 6 day lead time and for Z500 there are only small differences between IFS and IFS\_LR errors in our results in Fig.~\ref{fig:seasons}(c). 

As expected, IFS\_LR is ranked between ERA5 and IFS in terms of performance. Indeed, IFS and IFS\_LR forecasts are better than ERA5 forecasts because they start from the operational IFS analysis. Concerning PGW, which has the same horizontal resolution as IFS\_LR, it is interesting to note that both forecasts have similar errors to ERA5 for T850 over the winter period, and these are noticeably larger than with operational IFS forecasts. Also, the impact of the model horizontal resolution is smaller at longer lead times, as illustrated in Figs~\ref{fig:seasons}(c,d).

Finally, PGW appears performing better than IFS at day 2 for T850 and worse for Z500, particularly over the summer months (JJA) in Figs~\ref{fig:seasons}(a,b). Further investigations indicate that RMSE for Z500 and mean sea level pressure in PGW is much worse over the central Arctic than IFS. These results point to a fast-developing error over the central Arctic in that mass-field, that manifests at day 2 before other (non-systematic) errors start to dominate.

In Table~\ref{tab:fct}, we provide an overview of the forecasts analyzed in this study.
We also include the ensemble forecast run operationally at ECMWF (ENS) which is discussed in the next sections.
The 50-member ensemble has a horizontal grid spacing of 18km for the period considered here.
Please note, however, that the horizontal grid spacing of the ENS was recently reduced to 9km, consistent with the resolution of the IFS deterministic forecast \citep{lang2023}.

\begin{table}[t]
\centering
\small
\begin{tabular}
 {@{}c@{\hspace{\tabcolsep}}ll@{\hspace{\tabcolsep}}ll@{\hspace{\tabcolsep}}ll@{\hspace{\tabcolsep}}l@{}}
  \toprule
    Name & grid spacing & initialization & short description\\
  \midrule
    IFS  &  9km  &  operational IFS analysis & operational IFS high-resolution forecast\\
    PGW  &  28km &  operational IFS analysis & operation-like PanguWeather forecast\\
    IFS\_LR & 28km  & operational IFS analysis & IFS forecast run at a lower horizontal resolution\\
    PGW\_E5 & 28km  & ERA5 analysis   & PanguWeather forecast initialized with ERA5 \\
    ERA5 forecast & 28km  & ERA5 analysis & ERA5 hindcast (forecasts for a historical period) \\
    ENS (and EM)  & 18km  & ensemble operational analysis & operational IFS ensemble (end ensemble mean) forecast \\
\bottomrule
\\
\end{tabular}
\caption{List of forecasts investigated in this study.}
\label{tab:fct}
\end{table}

\section{Data and a case-study}
\label{sec:data}

We assess the performance of two upper-air variables, geopotential at 500 hPa (Z500) and temperature at 850 hPa (T850). Z500 and T850 forecasts are verified against the operational IFS analysis interpolated to a grid resolution of 1.5$^\circ$ following the World Meteorological Organisation (WMO) guideline and aggregated over the Northern Hemisphere. We also assess forecasts of 2m temperature against surface synoptic observations (SYNOP) over Europe. In addition, a verification of tropical cyclone (TC) forecasts is performed. Details about the verification process are provided in the Appendix.  

We show results mainly for two seasons: Summer 2022 (1 June 2022 to 31 August 2022) and  Winter 2022/2023 (1~December 2022 to 28~February 2023) to allow a focus on both extreme cold and extreme warm temperatures. These two verification periods are independent of the PGW training/validation dataset. Only results for Winter 2022 are shown for the upper variables because it is the most dynamically active season in the northern hemisphere. Following \cite{bi2022pangu}, the TC verification period covers 2 January to 30 November 2018. 

A comparison against SYNOP observations helps demonstrate the forecast performance from a user perspective. Nevertheless, in-situ observations have their drawbacks: the quality of the measurements is not perfect, the stations are not distributed homogeneously over the verification domain, and measurements can suffer discontinuity at a given station. Also, representativeness is a major concern when comparing model output with a point observation that might not be representative of the surrounding area. This representativeness issue is partially addressed here by an orography correction applied to the 2m temperature forecasts \citep[see for example][]{ingleby2015}. 

\begin{figure}[ht]
\begin{center}
\includegraphics[width = 0.95\textwidth,trim={0.cm 0.cm 0cm  0cm},clip]{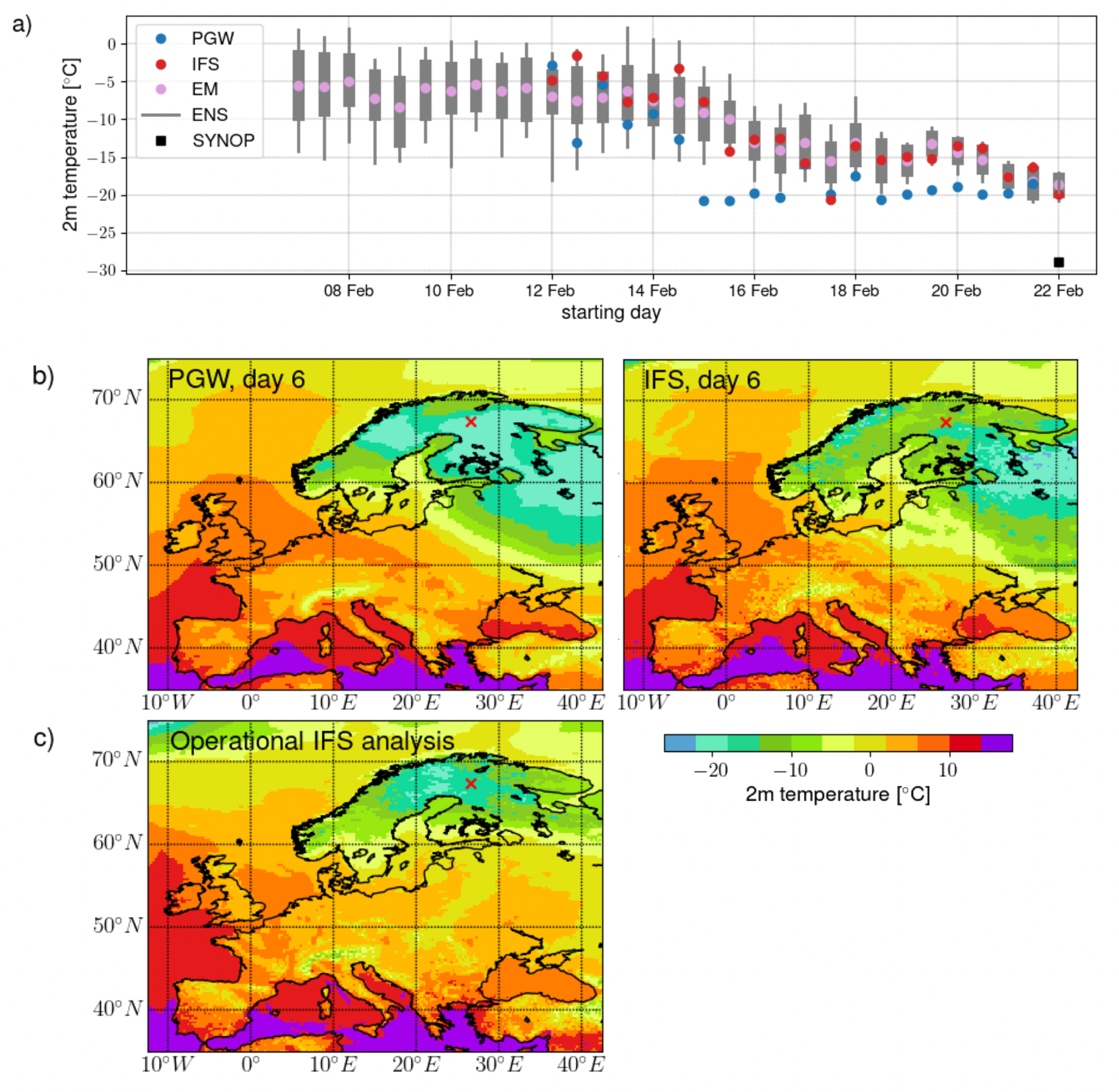}
\end{center}
\caption{
An example of 2m temperature forecasts and a corresponding SYNOP observation. (a) Evolution plots showing forecasts valid at Sodankyl{\"a}  (Finland) on 22 February 2023 at 00UTC: the 15-day ensemble forecast in the form of the ensemble mean and quantile forecasts (box-plots showing the 5\%, 25\%, 75\%, and 95\%), the 10-day PGW and IFS forecasts. (b) PGW (left) and IFS (right) forecasts at day 6, (c) verifying operational IFS analysis for Europe. The location of the Sodankyl{\"a}  SYNOP station indicated with a red cross on the maps.
}
\label{fig:case}
\end{figure}

An illustration of a forecast and observation is provided in Fig.~\ref{fig:case}. The forecast evolution over consecutive starting times shows that the ensemble spread becomes smaller as we approach the observation date. In Sondankyl{\"a} (Finland),  -29$^\circ$C was observed on that occasion. The PGW forecast has an earlier hint of the event severity than the IFS forecast, but both overestimated the temperature significantly to a similar degree. In the corresponding maps, PGW forecasts appear smoother than IFS forecasts, deprived of smaller scale structures. This first subjective assessment of a single case study agrees with the statistical analysis of the forecast performance discussed in the next section.

\section{Comparing forecast performance}
\label{sec:results}

\subsection{Contextualising the forecast skill}
\label{sec:quality}

\begin{figure}[]
\begin{center}
\includegraphics[width = 0.92\textwidth,trim={0cm 0cm 0cm 0cm},clip]{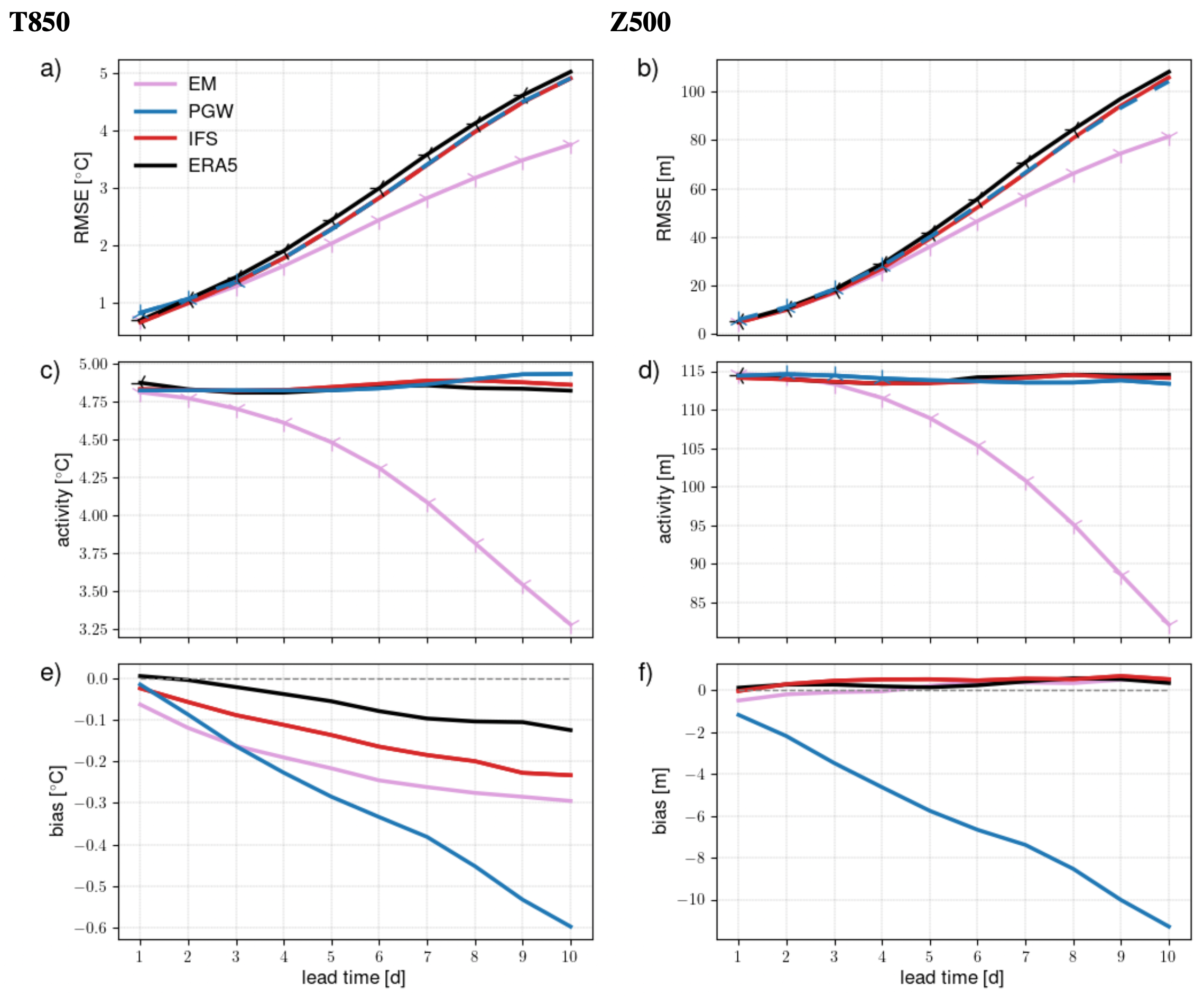}   
 \caption{(a,b) RMSE (the lower the better), (c,d) forecast activity, (the lower the forecast activity the smoother the forecast), and (e,f) forecast bias (the closer to zero the better), as a function of the forecast lead time for T850 (left panels) and Z500 (right panels). The forecasts are verified against the operational IFS analyses, and results are valid for Winter 2022/2023 over the Northern Hemisphere. A statistical significant difference with respect to the operational IFS forecast is indicated with a marker in (a), (b), (c) and (d). }
 \label{fig:t850}    
\end{center}
\end{figure}

Results in Fig.~\ref{fig:t850} (top row) are compelling: for lead times greater than 3 days, PGW forecasts are better than the ERA5 forecasts and as good as the operational IFS forecasts in terms of RMSE. The ensemble mean (EM), the ensemble functional that minimises the RMSE, is the best performing forecast with this metric.  RMSE is a key indicator of forecast performance but RMSE results need to be interpreted in light of other forecast characteristics that also contribute to the quality of a prediction. For example, in terms of realism, the ensemble mean cannot be considered as a plausible scenario of the atmospheric state.

In previous studies, there has been a concern that training towards RMSE results in overly smooth forecast fields \citep[see for example the smooth forecasts shown in][]{keisler2022forecasting}. Indeed, the RMSE strongly penalizes large forecast departures from the observations (or analyses), thus discouraging bold forecasts. When comparing RMSE from different models, it is therefore important to check the level of activity of the different forecasts while interpreting the results. The activity of a forecast is here defined as standard deviation of the forecast anomaly (see the Appendix for a formal definition). 
IFS and ERA5 forecasts have similar activity to each other and, importantly, similar activity to PGW for both T850 and Z500 (Fig.~\ref{fig:t850}, middle row). 

However, a clear smoothing of PGW forecasts at small scales is visible in Fig.~\ref{fig:case}. This dampening only has a minor contribution to the overall activity because this metric is dominated by larger scales.  We note that for Z500, the slight decrease in activity  with lead time of PGW is not statistically significant and in general PGW does not become smoother at longer lead times as confirmed by power spectra analysis (not shown).  This is not the case for the ensemble mean (EM), which becomes smoother as the forecast uncertainty increases. The lower activity of EM contributes to this good RMSE performance at longer time ranges. The unpredictable features are filtered out by averaging the ensemble members   
but forecast smoothness can be a non-desirable characteristic for some applications.  

Finally, Fig.~\ref{fig:t850} also compares the bias of the different forecasts. Ideally, a forecast should have a bias close to zero. The magnitude of the bias in PGW forecasts grows at a much faster rate than the bias in IFS or ERA5 forecasts, with the bias drift particularly strong for Z500. Whereas the bias of IFS and ERA5 stabilizes at longer lead times, the incremental bias in PGW forecasts is still present when extending the forecast horizon. 


\begin{figure}[]
\begin{center}    
\includegraphics[width = 0.95\textwidth]{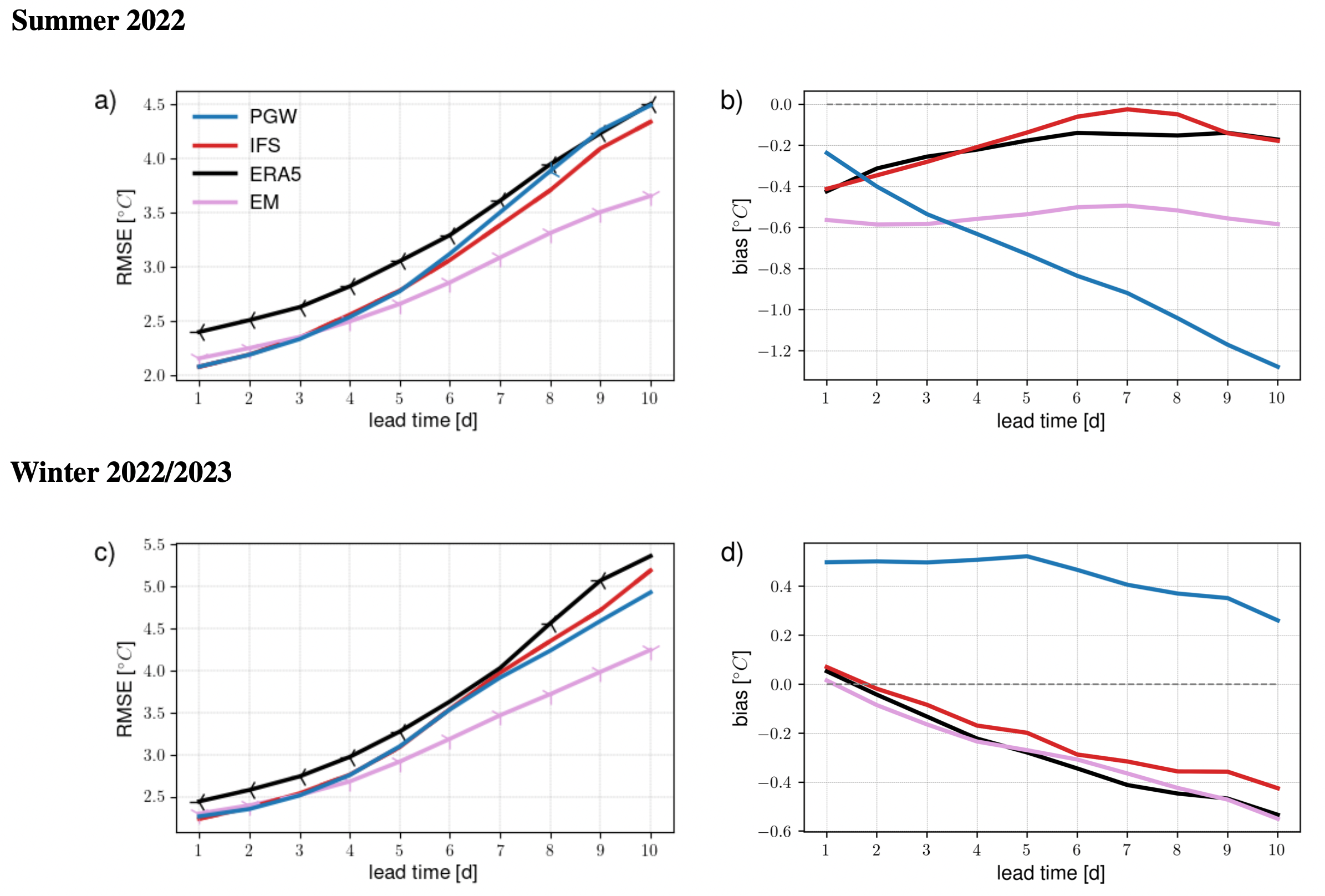}   
 \caption{Forecast performance for 2m temperature over Europe during Summer 2022 (top panels) and Winter 2022/2023 (bottom panels). (a,c) RMSE and (b,d) bias as a function of the forecast lead time. Summer 2022 forecasts are initialized at 12UTC (valid at midday) while Winter 2022/2023 forecasts are initialized at 00UTC (valid at midnight). The forecasts are verified against SYNOP observations. A statistical significant difference with respect to the operational IFS forecast is indicated with a marker in (a) and (c).}
\label{fig:surf}    
\end{center}
\end{figure}

Verification of upper variables against analysis is complemented by verification of 2m temperature against SYNOP observations in Fig.~\ref{fig:surf}. We find that verifying against observations shows similar results as against analysis for key metrics: the good performance of PGW forecasts in terms of RMSE, a bias drift for PGW with forecast lead time in summer, and EM outperforming the other forecasts from lead time day 4 onward, both in summer and in winter. Now, verification against observations is used as a framework for a more in-depth analysis of PGW forecast attributes. 

\subsection{Checking for statistical consistency}
\label{sec:consistency}

\begin{figure}[t]
\begin{center}    
\includegraphics[width = 0.95\textwidth]{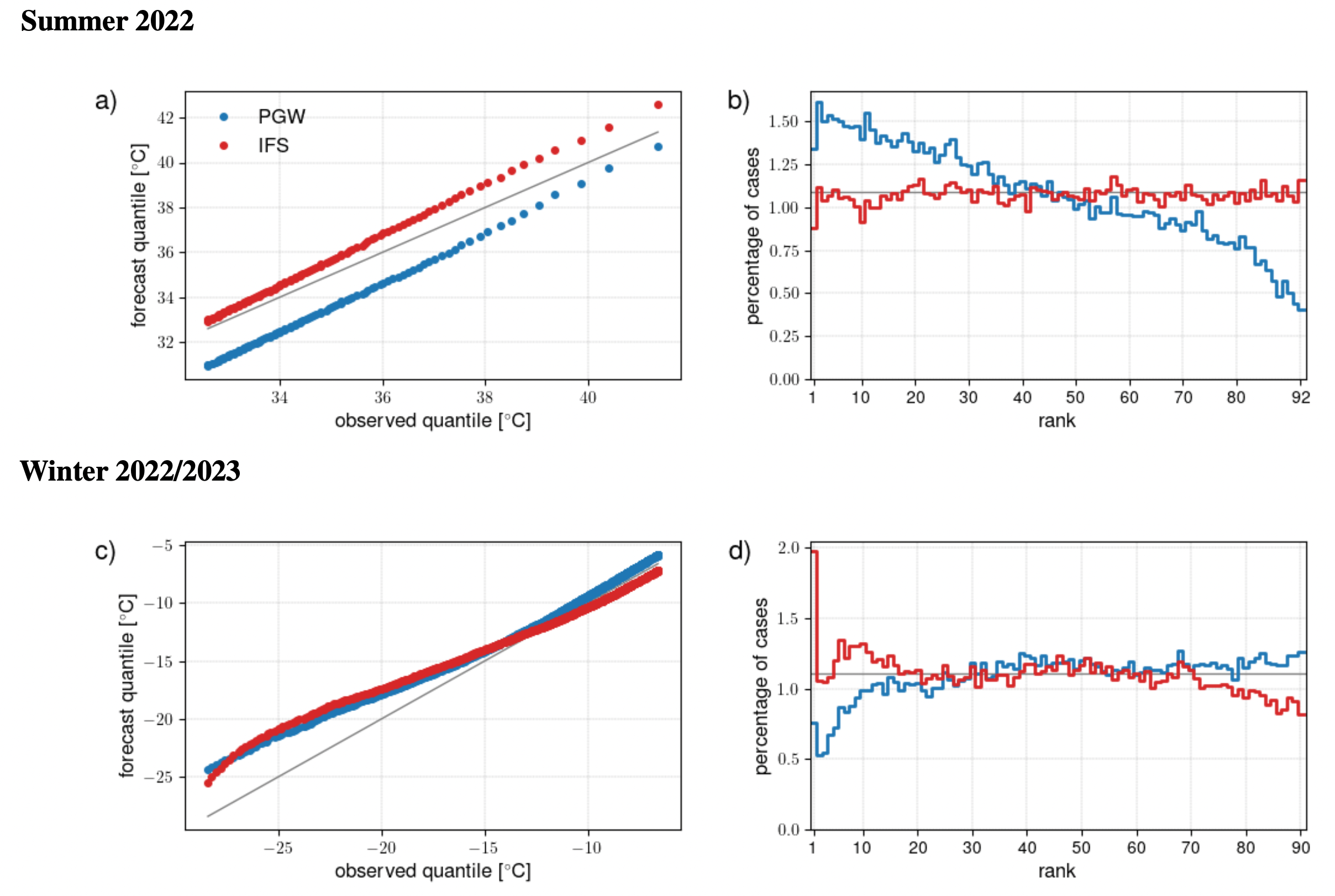}   
\caption{Statistical consistency of 2m temperature over Europe during Summer 2022 (top panels) and Winter 2022/2023 (bottom panels) for IFS and PGW forecasts at day 6. (a,c) Q-Q plots showing a scatter plot of the empirical forecast quantiles versus the quantiles from the observation distribution at quantile levels 90\%,90.1\%, …,99.9\% for the summer period (a)  and 0.1\%,0.2\%,..., 10.\% for the winter period (c). (b,d) Observation rank histograms show the averaged number of forecasts in the bins defined by the sorted observations at each station. For all plots, perfect reliability is indicated by a grey line.}
\label{fig:consistency}    
\end{center}
\end{figure}

We attempt now to assess the statistical consistency between the deterministic forecasts and the corresponding observations. Here we try to answer questions like ``Is the forecast able to mimic the observation statistical distribution?'', ``Is the forecast able to forecast extreme events of the same intensity as the observed ones?'' and ``Is the forecast systematically offset with respect to the observations?''.  Statistical consistency in terms of distribution is analysed using quantile-quantile (Q-Q) plots and observation rank histograms (a new type of diagnostic described in detail in the Appendix). The coherence of the spatial structures in the forecast would require additional diagnostic tools beyond the scope of this study. 

Q-Q plots for forecasts at day 6 focus on warm temperatures during summer in Fig.~\ref{fig:consistency}(a) and on cold temperatures during winter in Fig.~\ref{fig:consistency}(c).  For the former, both PGW and IFS forecasts can capture the observed extreme temperatures with PGW displaying a general offset consistent with the PGW bias at day 6 in Fig.~\ref{fig:surf}. For the latter, extremely low temperatures are not fully captured, neither by PGW nor by IFS, as already illustrated in the case study in Fig.~\ref{fig:case}.  In Northern Europe, very low temperatures are reached closest to the ground during clear-sky nights over snow-covered regions. This cooling is not fully captured by IFS during the evaluated period \citep{day2020}. 

Observation rank histograms (ORH) are used to check if the forecasts cover the observed range at each station separately. 
With OHR, we visualize how forecasts fall within the observed empirical distribution during the verification period. This diagnostic includes all stations rather than focusing on the hottest or coldest temperatures in the verification domain as in a Q-Q plot. A flat histogram indicates that the distribution of forecasts and observed temperatures is similar. This is the case for the IFS forecasts over the summer in Fig.~\ref{fig:consistency}(b) while the tilted histogram for PGW reflects a systematic bias in the forecast.  During winter, the IFS forecasts tend to be too cold at night time over mainland Europe \citep{sandu2020} while still not reaching extremely low temperatures in Northern Europe, as shown in Fig.~\ref{fig:consistency}. The overall negative bias leads to an over-populated first bin of the IFS histogram in Fig.~\ref{fig:consistency}(d) whilst PGW does not fully capture the lowest temperature at each station leading to underpopulated first bins of the histogram.

\subsection{Forecasting weather events}
\label{sec:events}

\begin{figure}[t]
\begin{center}    
\includegraphics[width = 0.95\textwidth]{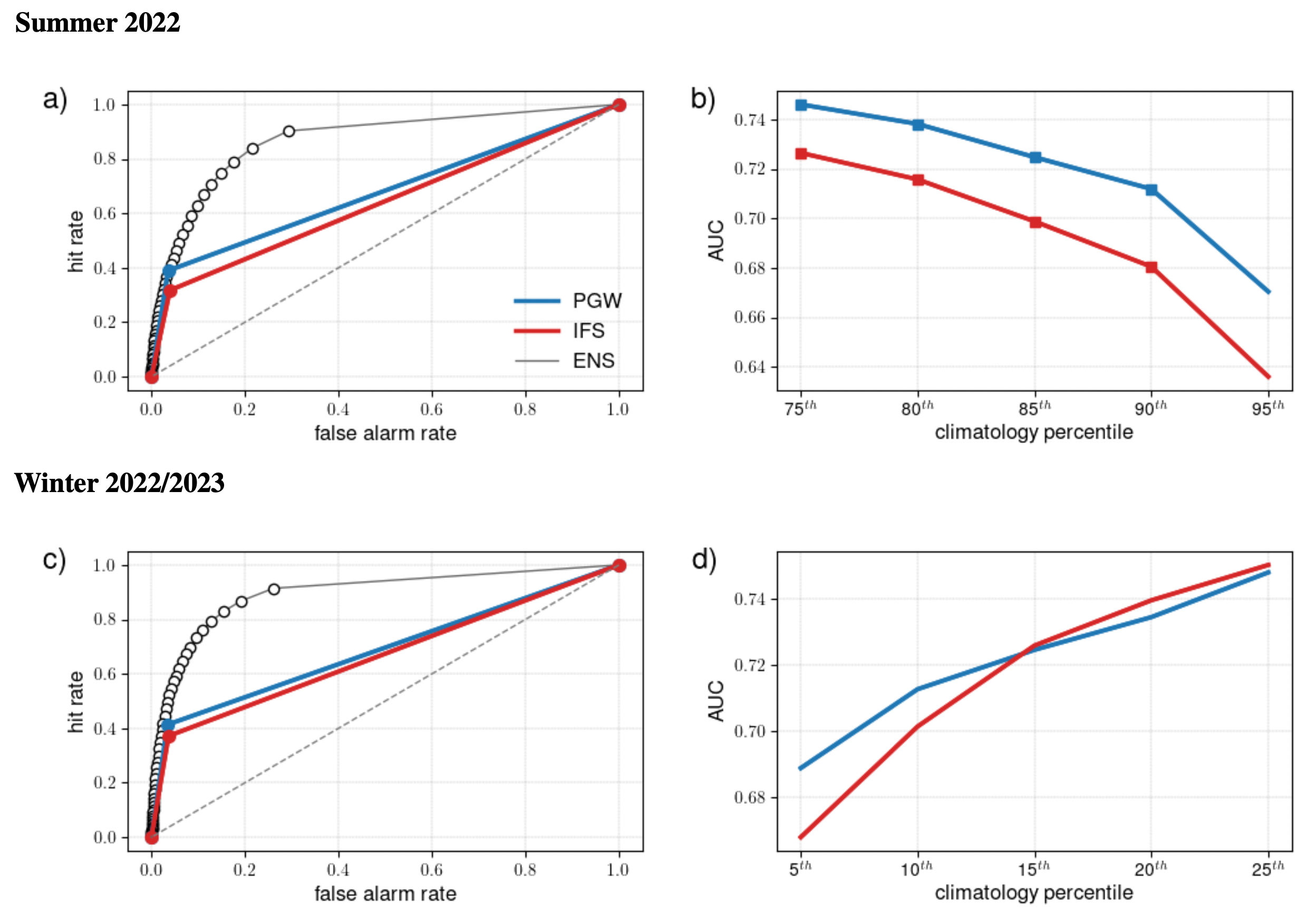}
\end{center} 
 \caption{Performance in forecasting 2m temperature events defined by climate thresholds for Summer 2022  (top panels) and Winter 2022/2023 (bottom panels) over Europe at day 6 lead time. 
 (a,c) ROC curve (the closer to the top left corner the better) for an event defined as exceeding the 95\% climate percentile in Summer (a) and below the 5\% climate percentile in Winter (c). The diagonal dashed line is the zero-discrimination line. Results for ENS-derived probability forecast are also shown. (b,d) Discrimination ability as measured with the area under the ROC curve (AUC, the higher the better) and plotted as a function of the climate percentile used to define a weather event. A statistically significant difference between PGW and IFS results (as estimated by block-bootstrapping) is indicated by a square. 
 }
    \label{fig:d_v}    
\end{figure}

The usefulness of a forecast is judged by its ability to predict weather events, often related to extremes. Here, the focus is on the forecast's ability to distinguish between an event and a non-event. Events are defined as 2m temperature exceeding a climate percentile. The climatology varies for each station and the climate percentiles are estimated based on the verification sample for the forecasts and the observations separately, to remove any bias in the forecast. We consider only low-temperature events for the winter period and high-temperature events for the summer period. 

The relative operating characteristic (ROC) curve is a popular diagnostic tool in forecast verification. ROC curves plotting hit rate versus false alarm rate of a high-temperature event in summer and a cold temperature event in winter are shown in Figs~\ref{fig:d_v}(a) and ~\ref{fig:d_v}(c), respectively.  
Deterministic forecasts such as PGW and IFS forecasts have only one non-trivial point on the curve.
This point is closer to the top-left corner of the plot for PGW than for IFS, indicating that PGW has better discrimination ability than IFS for the events under consideration. 
For the ensemble forecast, the ROC curve is built using one point for each probability issued by ENS using a standard `trapezoidal' approach  \citep{zbb2022}. The ROC curve of a probabilistic forecast, represented here by the empty circles, covers a much wider area than the curve derived from a single forecast.  

A standard measure for discrimination is the area under the ROC curve (AUC). In Figs~\ref{fig:d_v}(b,d), the AUC of 6-day ahead forecasts is plotted as a function of the percentile thresholds. The severity of the event increases as the climate percentiles get closer to 0\% in winter and 100\% in summer, indicating a rarer event under scrutiny.  
In general, AUC decreases when focusing on more intense/rare events as it becomes more difficult to predict such events with a deterministic forecast. IFS and PGW have similar levels of performance in winter while PGW outperforms IFS in summer with differences statistically significant for percentiles between 75\% and 90\%, as estimated by block-bootstrapping. 
Combining results in Figs~\ref{fig:consistency} and \ref{fig:d_v}, we see that PGW climatology for summer extremes is less consistent with the analysis climatology, but after accounting for this discrepancy, PGW has more accurate forecasts of summer temperature extremes than IFS.

\subsection{Forecasting tropical cyclones}
\label{sec:tcs}

Tropical cyclones (TCs) are a prominent example of extreme weather that has a devastating impact and attracts considerable attention from the public and media. Moreover, TCs are characterized by large deviations from the mean state of the atmosphere and are thus generally challenging to forecast.  Here, we focus on the year 2018 \citep[as is done in][]{bi2022pangu}, but note that IFS TC forecasts have substantially improved with more recent cycles \citep{forbes2021,majumdar2023}. We assess 2 key characteristics of TCs: their track position and their intensity (see the Appendix for more details). In Fig.~\ref{fig:tc}(a), the position error is measured as the distance between the TC position in the forecasts and the observations at a specific time. Larger errors are observed for PGW during the first day compared with IFS, but PGW has slightly lower errors for lead times greater than 2 days. This difference is partly explained by the fact that the propagation speed is generally too slow in the IFS \citep{chen2023} but not in PGW (not shown). Overall, the differences in position error are small and not statistically significant between models. 

Focusing now on TC intensity, Fig.~\ref{fig:tc}(b) shows the mean absolute error for TC central pressure. Here we find that PGW clearly underestimates the intensity (\textit{i.e.} the predicted pressure is too high). Both IFS and  ERA5 perform better than PGW in terms of TC intensity error (except for at 0 day lead time). The large positive bias of PGW in the minimum core pressure results from too-weak gradients and too-weak maximum wind speed, while the IFS more closely resembles the analysis (not shown). The better performance of IFS compared with ERA5 is mainly explained by its higher resolution (9 km vs 28 km) but is also due to improvements in IFS through model development. 

The number of unique TCs observed in our verification dataset is 107. We counted 105 (86) TCs in IFS and 95 (61) in Pangu at the initial time (day 5). Among the observed TCs, 51 are of category 1 or above.  We count 49 (34) TCs of these categories in IFS and 46 (33) in Pangu at the initial time (day 5). So the difference in the number of TC predicted between IFS and PGW appears to be a function of the TC intensity. Further investigations are needed to explore the root causes for the reduced number of low-intensity TCs in the ML-based forecast as well as to closely examine the TC structures and physical consistency between variables. 

\begin{figure}[]
\begin{center}
\includegraphics[width = 0.9\textwidth,trim={0cm 0.cm 0cm 0cm},clip]{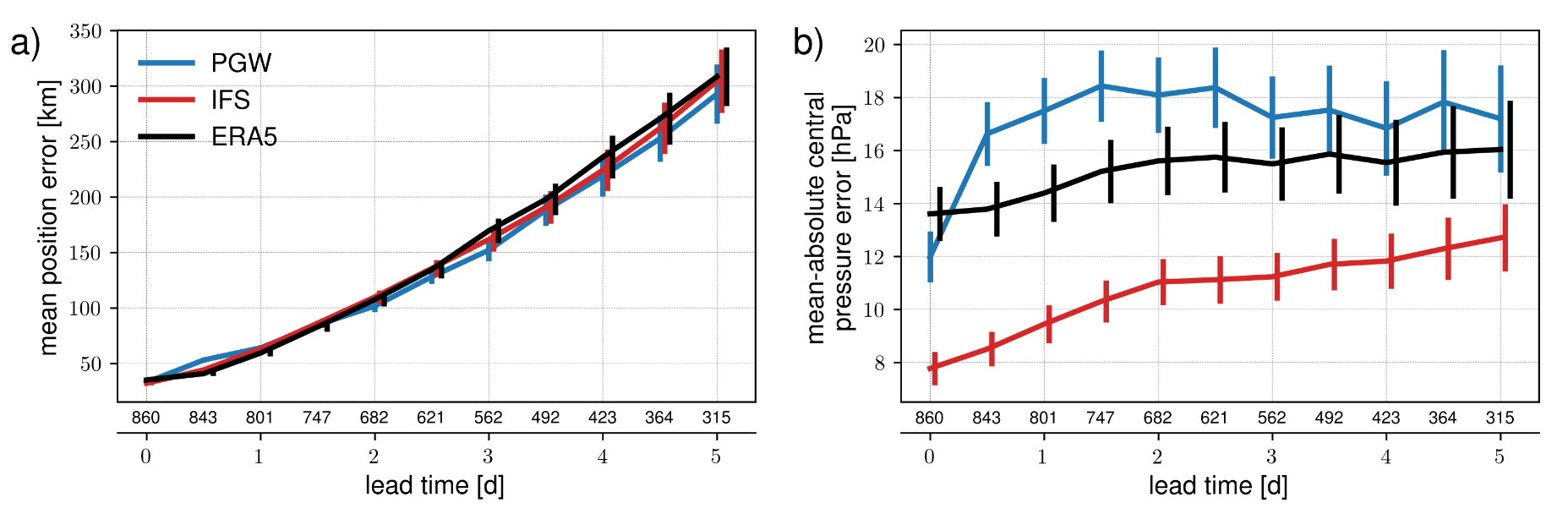} 
\caption{Tropical cyclone verification results: (a) mean position error and (b) mean absolute central pressure error as a function of the lead time for 2018. Forecasts are verified against the IBTrACS dataset and homogenized to have a consistent number of cases between models. For each lead time, the number of cases is displayed directly below the graphs. The vertical bars indicate the 2.5\%-97.5\% confidence intervals. }
\label{fig:tc}    
\end{center}
\end{figure}

\subsection{Predicting the forecast error}
\label{sec:predictability}

The day-to-day variability of the error is compared for PGW and IFS forecasts. We aim to identify common patterns in error growth and examine the sensitivity to predictability barriers. For this purpose, we analyze the so-called \textit{predictability barrier plots}: 2D diagrams displaying the forecast error as a function of both the forecast starting time (x-axis) and the forecast lead time (y-axis). A more in-depth analysis would involve running different models from different initial conditions as in \cite{magnusson2019} but this approach is out of scope for this paper. 

\begin{figure}[ht]
\begin{center}
\includegraphics[width = 0.89\textwidth,trim={0cm 0cm 0cm 0cm},clip]{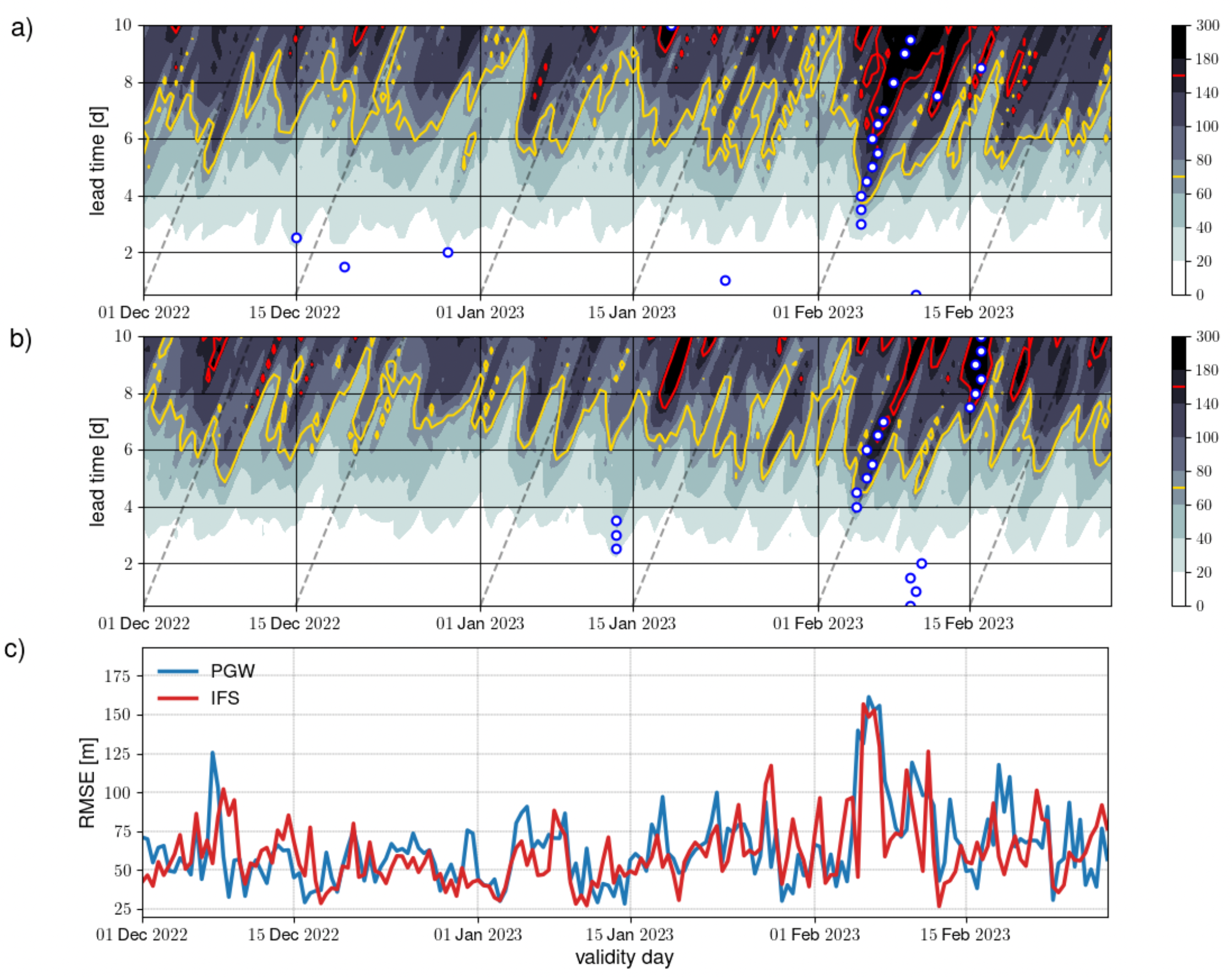}    
\caption{Predictability barrier plots showing daily RMSE for Z500  for lead times 1 to 10 days over Winter 2022/2023 for PGW (a) and IFS (b). A cross-section at day 6 of Figs~(a) and (b) is provided in (c). In (a) and (b), the shade indicates the score value (in m), the vertical lines intercept scores for all forecasts valid on a given day, the transversal lines intercept scores for a given forecast run for all lead times, the yellow lines indicate the averaged score for a day 6 forecast, the red lines mark a large error, and the blue dots indicate the worst score over the period for each lead time.}\label{fig:predictability}    
\end{center}
\end{figure}

Examples of predictability barrier plots for PGW and IFS are provided in Figs~\ref{fig:predictability}(a) and \ref{fig:predictability}(b), respectively, focusing on daily scores of Z500 forecasts over Europe. In these plots, a transverse structure indicates rapid error growth leading to a poor forecast at all lead times:  in that case, the forecast initialization might be the dominant predictability limiting factor. By contrast, a vertical structure indicates a weather situation difficult to predict for consecutive runs with different initialization, likely to be due to predictability barriers for that specific weather situation.

In general, we see a good agreement between PGW and IFS daily errors. This similarity is even more evident when plotting daily errors for a single lead time (here day 6) for the whole verification period. The correlation coefficient between the two time-series is 0.54. Strikingly, Fig.~\ref{fig:predictability}(c) shows the same `bust' in forecasting the weather over Europe for 6$^\text{th}$ February. This flow-dependent nature of the error points towards the need for ensemble forecasting in a similar fashion for ML models like PGW as is common practice for NWP nowadays.

\section{Summary and outlook}
\label{sec:summary}

The results shown here highlight that machine learning (ML) models could have a promising future in numerical weather prediction. To explore the advantages and limitations of data-driven weather forecasts, we have run PanguWeather, an ML model trained on ERA5, initialized with the operational IFS analysis. The PGW forecasts are compared with the operational IFS forecasts to help shape our understanding of the characteristics of both the data-driven forecasts and their errors. Some of the most challenging weather phenomena are linked to rain, but our comparison does not include precipitation because the field is not present in PanguWeather.

Fundamentally, data-driven forecasts show good performance with the ML model being skillful for both upper-air variables (geo-potential height at 500hPa and temperature at 850hPa) when verified against operational IFS analysis and for a surface variable (2m temperature) against observations. These conclusions are supported by further investigations including for other variables such as 10m wind speed. These results are not shown here because they are in line with the findings of this study.

We note, however, that there is room for further improvement in the implementation of data-driven weather prediction systems. Like all data-driven approaches presented so far, the ML model used in our study is not trained on the operational IFS analysis and our experiments do not involve fine-tuning. Instead, ERA5 is used as the training dataset, which has so far been the cornerstone of any data-driven approach. For some aspects, ML models appear to directly inherit advantages and drawbacks from the numerical weather prediction system used to generate the training dataset. For example, the model resolution of the training data can in part explain the limitation in forecasting small-scale structures with PanguWeather. However, the forecast initial conditions also play a crucial role: starting a forecast from the operational IFS analysis rather than ERA5 analysis offers an advantage in accuracy also for the medium range. Moreover, the similarities in error growth of a data-driven forecast and a standard NWP forecast indicate similar sensitivities to chaos between ML-based and physically-based models.  

The data-driven forecast appears smoother than the operational IFS forecast but the level of smoothness does not seem to increase with the forecast lead time, as we might expect when training toward RMSE. However, we observe a drift in bias almost linear with the forecast lead time. While this drift could be addressed when developing future ML models, statistical post-processing offers a means to correct systematic errors and achieve improved forecasts \citep{vannitsem2021}. A deeper understanding of systematic errors could be achieved by performing conditional verification, for example, focusing on specific physical processes. Moreover, other diagnostic tools could be used to check for physical consistency based, for instance, on multivariate verification that accounts for the correlation between variables. 

Good performance of data-driven forecasts is also observed in predicting some extreme events and confirmed by case studies. The results shown here focus first on events defined as climate threshold exceeded at a station location. The performance of ML-based models in forecasting TCs is under scrutiny too, as in \cite{bi2022pangu}. 
Preliminary investigations indicate that current ML models, due to their lower resolution, tend to predict less weak TCs compared to IFS, with tracks of similar quality, but IFS better captures their intensity and structure.
Additional studies would help to demonstrate the value of data-driven forecasts as well as their strengths and weaknesses in supporting decision-making.

Finally, in this work, we focused on deterministic forecasts but ensemble forecasts are key in providing uncertainty information for decision-making. A Monte-Carlo approach for uncertainty quantification has been tested starting ML-derived forecasts from perturbed initial conditions based on the ECMWF ensemble data assimilation and singular vector perturbations. The initial condition perturbation methodology is described in \cite{enspert}. The resulting ensemble forecast is showing promising results. In future work, uncertainty in initial conditions will be complemented by mechanisms to account for model uncertainty \citep[see e.g.][]{Lang21} in a data-driven weather prediction context.

This first assessment of a machine learning-based weather forecast in an operational-like context shows very promising results. The future role of ML models in the context of numerical weather prediction systems, and the ability of this approach to complement physical models remains to be explored. Operational centres should explore the strengths and weaknesses of these models as additional components of their forecasting systems: the ability to run forecasts at a much higher speed and much lower computational cost opens new horizons. 

\bibliographystyle{ametsocV6}

\section*{Appendix: verification process and scores definition}
\label{sec:eval}

\subsection*{Contextualising the forecast skill}
We aim at a quantitative comparison of PGW and IFS forecasts. In general terms, forecast verification consists of measuring the relationship between a forecast and the corresponding observation\footnote{the term \textit{observation} is used in the broad sense of an assumed `truth'. It can take the form of an analysis or an observation.}  \citep{murphy87}. For this purpose, one can carefully choose from a variety of metrics and diagnostics  \citep[see][]{wilksv1,jolliffe}. To go beyond the computation of generic scores, it is possible to investigate the properties of the joint distribution of forecasts and observations. The two main forecast attributes are forecast consistency (or calibration) and forecast discrimination ability. Note that these concepts hold also when dealing with probabilistic forecasts (which are not explored here).

Classical statistic tools involve the computation of summary statistics and scores. Summary statistics include the bias defined as the averaged difference between forecasts and observations, and forecast/observation activity defined as the standard deviation of the forecast/observation anomaly. Scores are metrics measuring the forecast accuracy such as the anomaly correlation shown in Fig.~\ref{fig:hsc} or the forecast error such as the widely-used RMSE shown in Fig.~\ref{fig:seasons}. A formal definition of each of these metrics is provided in the Appendix. 
Here we formally define the following quantities:
\begin{itemize}
\item the forecast root mean squared error:
\begin{equation}
\sqrt{\overline{(f-o)^2}},
\end{equation}
\item the forecast mean error (or bias):
\begin{equation}
\overline{f-o},
\end{equation}
\item the forecast activity:
\begin{equation}
\sqrt{\overline{\left[(f-c)-\overline{f-c}\right]^2}},
\end{equation}
\item the observation activity:
\begin{equation}
\sqrt{\overline{\left[(o-c)-\overline{o-c}\right]^2}},
\end{equation}
\item the forecast anomaly correlation:
\begin{equation}
\frac{\overline{(f-c-\overline{f-c})(o-c-\overline{o-c})}}{\sqrt{\overline{(f-c-\overline{f-c})^2}}\sqrt{\overline{(o-c-\overline{o-c})^2}}}    
\end{equation}
\end{itemize}
with $f$ the forecast, $o$ the observation, $c$ the climatology, and $\overline{\cdot}$ the averaging operator including a latitude weighting.

\subsection*{Checking for statistical consistency}
Statistical consistency is tested regionally with quantile-quantile (Q-Q) plots and locally (at the station level) with observation rank histograms. For this exercise, we exclude stations situated at an altitude greater than 1000m to avoid focusing predominantly on representativeness issues rather than model characteristics. For Q-Q plots, quantiles are estimated from the whole verification sample (Europe, Summer 2022 or Winter 2022/2023) for both observations and forecasts, separately. We restrict our analysis to the warm tail of the distribution in the summer (quantile levels in the range $90\%-99.9\%$) and to the cold tail of the distribution in the winter  (quantile levels in the range $0.1\%-10\%$).  

As a complementary diagnostic tool, we suggest a new type of plot: the observation rank histogram (ORH). Inspired by the ensemble rank histogram used to assess the reliability of ensemble forecasts, ORH is built by ranking the observations from the smallest to the biggest for the whole verification period and individual forecasts, for each station separately. The rank of the forecast for each verification day is registered and populates the histogram. ORH assesses whether forecasts and observations are distributed similarly at a station level.  

\subsection*{Forecasting weather events}
Forecasting specific events is at the heart of many weather applications. The ability of a forecast to distinguish between the occurrence and non-occurrence of an event is called discrimination. Based on local climatology, event thresholds are defined as discussed above. Forecasts and observations are transformed into binary values with respect to a given threshold. A contingency table is populated for each of the dichotomous events. A contingency table is a $2\times2$ table where hits, misses, false alarms, and correct negatives are counted. From this table, it is possible to derive both the hit rate and the false alarm rate, the two components of the relative operating characteristic (ROC) curve. The area under the ROC curve (AUC) is a common measure of discrimination in weather forecast verification \citep{mason82,harvey92}. 

Weather events are defined with the help of a local climatology that differs for each station. The same verification setting is used as in \cite{zbb19} where an event is defined using a percentile of a climatology rather than a fixed absolute value. This approach tries to reflect that user-relevant thresholds are often associated with potential hazards and as such vary from place to place. For example, the 5\% percentile of the local temperature climatology corresponds to very different absolute thresholds for say Helsinki and Madrid. Also using a climatology-based threshold allows us to avoid the pitfall of measuring varying climatology rather than actual skill \citep{hamilljuras2006}. 

A different climatology is defined for (i) the observations and (ii) each forecast, with percentiles directly estimated from the verification sample. This so-called \textit{eigen-climatology} approach corresponds to practically applying an in-sample local bias correction of the forecast as discussed in more detail in \cite{zbb19}. This step is important to disentangle discrimination from calibration attributes, because the latter can, in principle, be improved by post-processing.  Only stations where measurements are available throughout the full verification period are considered for this exercise.

Finally, we recall that statistical significance is important when comparing competing forecasts \citep{geer2016}. Here, we assess the chaotic variability of the scores with the use of (block)-bootstrapping. We randomly choose the verification days entering the verification dataset and compute scores for each forecast. Based on a 1000-member block bootstrap sample with blocks of 5 days, statistical significance to the 5\% level is estimated. 

\subsection*{Forecasting tropical cyclones}
We also assess performance in forecasting tropical cyclones (TCs). TCs are tracked in forecasts from PGW, IFS, and ERA5 with the ECMWF operational TC tracker as described in \cite{magnusson2021}. Forecasts up to 5 days are verified here as results for longer lead times are unlikely to be statistically significant. As observations, we use the International Best Track Archive for Climate Stewardship (IBTrACS) database \citep{knapp2010,knapp2018}. The verification is based on TCs that are present in the observation database at the forecast initial time. The sample is homogenized to include the same cases for all 3 models. This homogenization results in a sample size of 860 cases at the analysis time, down to 315 cases at day 5\footnote{If only the IFS was validated, the maximum number of cases would be 988 at the forecast initial time and 592 for 5-day forecasts.}. An intensity threshold of 17m/s is applied to filter the observation dataset for TCs that reach tropical storm strength.

\end{document}